\documentclass[a4paper]{article}
\usepackage{fullpage}

\usepackage{booktabs}
\usepackage{adjustbox}
\usepackage[group-separator={,}]{siunitx}
\usepackage{wrapfig}
\usepackage{enumitem}

\setlist[itemize]{noitemsep,topsep=0pt}

	\usepackage{ruby} 
	\newcounter{Rubycount}
\makeatletter
\newcommand{\myLabel}[2]{#2
\protected@write \@auxout {}{\string \newlabel {#1}{{#2}{}}}}
\makeatother

\newcommand{\Phoni}[1]{\ensuremath{\mathsf{PHONI}_{#1}}}
\newcommand{\Moni}{\ensuremath{\mathsf{MONI}}}

\newcommand*{\PlusPlus}{\kern0.3ex\raisebox{-0ex}{\scalebox{0.8}{\kern-0.4ex+}}\kern-0ex\raisebox{0.5ex}{\scalebox{0.8}{\kern-0.4ex+}}}
\newcommand*{\CPlusPlus}{C\PlusPlus{}}

\usepackage{tikz}
\usepackage{pgfplots}
\usepgfplotslibrary{colorbrewer}

\pgfplotsset{cycle list/Dark2,
cycle multiindex* list={mark list*\nextlist
Dark2\nextlist
},
}

\pgfplotsset{width=150mm,height=100mm,
major grid style={draw=none},
minor grid style={draw=none},
grid,
every axis/.append style={line width=0.5pt,
tick style={line cap=round,
thin,
major tick length=4pt,
minor tick length=2pt,
},
},
legend cell align=left,
legend pos=north west,
}

\pgfplotsset{squeezedPlot/.style={width=0.4\linewidth,
height=0.3\linewidth
},
appendLegend/.style={transpose legend=true,
legend style={font=\small}}
}

\usepackage{epsfig}
\usepackage{amsmath}
\usepackage{amssymb}
\usepackage{color}
\usepackage{url}

\newlength{\figurewidth}
\newlength{\smallfigurewidth}
\usepackage[ruled,vlined,linesnumbered]{algorithm2e}
\SetKwComment{Comment}{$\triangleright$\ }{}

\usepackage{amsthm}
\theoremstyle{definition}
\newtheorem{theorem}{Theorem}
\newtheorem{definition}[theorem]{Definition}

\newcommand{\bwt}{\ensuremath{\mathsf{BWT}}}
\newcommand{\isa}{\ensuremath{\mathsf{ISA}}}
\newcommand{\lce}{\ensuremath{\mathsf{LCE}}}
\newcommand{\lcp}{\ensuremath{\mathsf{LCP}}}
\newcommand{\lf}{\ensuremath{\mathsf{LF}}}
\newcommand{\ms}{\ensuremath{\mathsf{MS}}}
\newcommand{\NULL}{\ensuremath{\mathsf{NULL}}}
\newcommand{\plcp}{\ensuremath{\mathsf{PLCP}}}
\newcommand{\rank}{\ensuremath{\mathsf{rank}}}
\newcommand{\sa}{\ensuremath{\mathsf{SA}}}
\newcommand{\select}{\ensuremath{\mathsf{select}}}
\newcommand{\pos}{\ensuremath{\mathsf{pos}}}
\newcommand{\len}{\ensuremath{\mathsf{len}}}
\newcommand{\ignore}[1]{}
\newcommand{\Pos}{\pos}
\newcommand{\Len}{\len}

\begin{document}

\title{\large \textbf{PHONI: Streamed Matching Statistics\\with Multi-Genome References}\thanks{MR, TG, BL and CB are funded National Science Foundation NSF IIBR (Grant No. 2029552) and National Institutes of Health (NIH) NIAID (Grant No. HG011392).  CB is funded by NSF SCH: INT: (Grant No. 2013998). 
MR and CB are funded by NSF IIS (Grant No. 1618814) and NIH NIAID (Grant No. R01AI141810).  
TG is funded by NSERC Discovery Grant RGPIN-07185-2020.  
DK is funded by JSPS KAKENHI Grant JP18F18120.
GN funded by Basal Funds FB0001 and Fondecyt Grant 1-200038, ANID, Chile. AP funded by Basal Funds FB0001 and Doctoral Scholarship grant 21180760, ANID, Chile.}
}

\date{}

\author{Christina Boucher$^{\ast}$, Travis Gagie$^{\dag}$, Tomohiro I$^{\ddag}$, Dominik K\"oppl$^{\S}$,\\
Ben Langmead$^{\P}$, Giovanni Manzini$^{\parallel}$, Gonzalo Navarro$^{\ast \ast}$,\\ Alejandro Pacheco$^{\ast \ast}$ and Massimiliano Rossi$^{\ast}$\\[0.5em]
{\small\begin{minipage}{\linewidth}\begin{center}
\begin{tabular}{cccc}
$^{\ast}$U Florida & $^{\dag}$Dalhousie U & $^{\ddag}$Kyutech & $^{\S}$ TMDU\\
Gainesville, USA & Halifax, Canada & Fukuoka, Japan & Tokyo, Japan\\
\url{{cboucher,rossi}} & \url{travis.gagie} & \url{tomohiro} & \url{koeppl.dsc}\\
\url{@cise.ufl.edu} & \url{@dal.ca} & \url{@ai.kyutech.ac.jp} & \url{@tmd.ac.jp}\\
&&& (corresponding)
\end{tabular}\\[2ex]
\begin{tabular}{ccc}
$^{\P}$Johns Hopkins U & $^{\parallel}$U Piemonte Orientale & $^{\ast \ast}$CeBiB, DCC, U Chile\\
Baltimore, USA & Alessandria, Italy & Santiago, Chile\\
\url{langmea} & \url{giovanni.manzini} & \url{{gnavarro,alpachec}}\\
\url{@cs.jhu.edu} & \url{@uniupo.it} &  \url{@dcc.uchile.cl}
\end{tabular}
\bigskip
\end{center}\end{minipage}}}

\maketitle
\thispagestyle{empty}

\begin{abstract}
	Computing the matching statistics of patterns with respect to a text is a fundamental task in bioinformatics, but a formidable one when the text is a highly compressed genomic database.  Bannai et al.\ gave an efficient solution for this case, which Rossi et al.\ recently implemented, but it uses two passes over the patterns and buffers a pointer for each character during the first pass.  In this paper, we simplify their solution and make it streaming, at the cost of slowing it down slightly.  This means that, first, we can compute the matching statistics of several long patterns (such as whole human chromosomes) in parallel while still using a reasonable amount of RAM; second, we can compute matching statistics online with low latency and thus quickly recognize when a pattern becomes incompressible relative to the database.  Our code is available at \mbox{\url{https://github.com/koeppl/phoni}}~.
\end{abstract}

\section{Introduction}
\label{sec:introduction}

Computing the matching statistics of patterns with respect to a text is a fundamental and well-studied task in bioinformatics, useful in many applications~\cite{MBCT15,Ohl13}, since they tell us which substrings of those patterns occur in that text.  We consider a slightly extended definition, which includes some information on where those substrings occur.

\begin{definition}
\label{def:matching_statistics}
The {\em matching statistics} \ms\ of a pattern $P [0..m - 1]$ with respect to a text $T [0..n - 1]$ are an array of (position, length)-pairs $\ms [0..m - 1]$ such that
\begin{itemize}
\item $P [i..i + \ms [i].\len - 1] = T [\ms [i].\pos..\ms [i].\pos + \ms [i].\len - 1]$,
\item $P [i..i + \ms [i].\len]$ does not occur in $T$.
\end{itemize}
That is, $\ms [i].\pos$ is a pointer to the starting position of a copy in $T$ of the longest prefix of $P [i..m - 1]$ that occurs in $T$, and $\ms [i].\len$ is the length of that prefix.
\end{definition}

For example, if $T [0..5] = \mathtt{CATTAG}$ and $P [0..4] = \mathtt{GTTAC}$, then $\ms [0..4]$ can be either $[(5, 1), (2, 3),$ $(3, 2), (4, 1), (0, 1)]$ or $[(5, 1), (2, 3), (3, 2), (1, 1), (0, 1)]$. The pair $\ms [1] = (2, 3)$ tells us that the prefix $P [1..3] = \mathtt{TTA}$ of $P [1..4]$ occurs in $T$ at $T [2..4]$ and $P [1..4]$ does not occur in $T$.

Despite the importance of computing matching statistics, until recently it was not known how to index efficiently a compressed representation of a massive and highly repetitive text, such as a database of genomes of individuals from the same species, so that later we could quickly compute the matching statistics of a given pattern.  Bannai et al.~\cite{bannai2020refining} augmented Gagie et al.'s~\cite{gagie2020fully} r-index to support computation of matching statistics, as we will describe in Section~\ref{sec:simplifying}, but they did not say how to build their auxiliary data structure.  Rossi et al.~\cite{MONI} recently gave a construction, implemented it, and used it to find maximal exact matches (MEMs) between a set of DNA reads and a genomic database.  They called their implementation MONI, Finnish for ``multi'', since their ultimate goal is to align reads to a multi-genome reference.

Bannai et al.'s solution makes two passes over the pattern: during the first, it works from right to left and computes and buffers the $\pos$ values; during the second, it works from left to right and uses the $\pos$ values and random access to the text to compute the $\len$ values.  This means the working space grows linearly with the length of the pattern.  This is not a serious concern when the patterns are short reads or even long reads, which are generally hundreds or thousands of characters, respectively, but it could limit how many extremely long patterns we can process in parallel.

A human chromosome~1 is a quarter of a billion base pairs, so assuming a $\pos$ values takes 8 bytes, we could need 2 GB of RAM to buffer the $\pos$ values, on top of the RAM occupied by Bannai et al.'s index.  If we have dozens of cores and want to process a chromosome 1 from a different haplotype on each core in parallel, the RAM needed to buffer the $\pos$ values could easily dwarf that needed for the index.  Bannai et al.\ themselves proposed computing the matching statistics of whole genomes with respect to genomic databases, in order to identify regions of novel DNA --- where the $\len$ values are small, meaning the region is incompressible relative to the database --- for rare-disease diagnosis.  Such whole-genome matching statistics could also be useful in estimating genetic diversity in a population, tracking how pathogens mutate, etc.

Another potential concern with Bannai et al.'s solution is that, if a pattern is being given to us online, character by character, then the maximum latency, from the time we receive a character to the time we return the corresponding pair in the matching statistics, also grows linearly with the length of the pattern.  This means we cannot use Bannai et al.'s solution for applications in which we want the matching statistics in real time.

On the other hand, if we can stream the matching statistics in real time then, among other things, we can use the results in applications of DNA sequencing that require rapid computational feedback.  For example, when the sequencing process in Oxford Nanopore MinION DNA sequencers starts to degrade in accuracy, their output continues but it no longer contains useful information~\cite{oliva2020portable}.  With streamed matching statistics, we can stop the sequencing process when the output becomes incompressible relative to the database.  Even when the MinION is still producing valid output, it may be sequencing DNA that is irrelevant for our purposes.  Real-time computation of matching statistics also allows us to reject such DNA rapidly, and thus target the sequencer to specific genomes or genes of interest~\cite{edwards2019real,kovaka2020targeted}.  Finally, we may want to stop sequencing once the MinION has found a long enough match to confirm the presence of a pathogen, for example~\cite{taxt2020rapid}.  For all these applications, reducing latency optimizes throughput, and reducing the memory usage allows decisions to be made ``close to'' the sequencer using embedded or other non-server processors.

In this paper, we simplify Bannai et al.'s solution by using longest-common-extension (LCE) queries to compute the $\len$ values at the same time that we compute the $\pos$ values.  This means we process patterns using a single pass, reading them and writing the matching statistics as streams, without buffering --- so our solution can be applied to the tasks described above.  (To stream patterns from left to right we should really index the reversed texts, but we ignore that in this version of this paper.)
Our experiments show that our implementation, which we call PHONI, needs significantly less time and memory to build than MONI; is significantly smaller once built; and uses less extra RAM to compute the matching statistics of long patterns.  For the full version of this paper we will test the maximum latency and the RAM usage when the solutions process queries in parallel.

The rest of this paper is laid out as follows: in Section~\ref{sec:review} we review how to compute and use \ms{}, data structures supporting random access to $T$, and data structures supporting \lce\ queries; in Section~\ref{sec:simplifying} we explain how to simplify Bannai et al.'s algorithm to perform the computation within a single pass by means of \lce\ queries; and in Section~\ref{sec:experiments} we present our experimental results.  
Our code is available at \url{https://github.com/koeppl/phoni}\ .

\section{\ms, Random Access and \lce}
\label{sec:review}

\subsection{Computing and Using \ms}
\label{subsec:ms}

There are practical $O (n \log \sigma)$-bit data structures with which we can compute \ms\ in $O (m \log \sigma)$ time, where $\sigma$ is the size of the alphabet of $T$~\cite{belazzougui2018fast}.  Once we have \ms\ we can easily compute in $O (m)$ time the maximal exact matches (MEMs) of $P$ with respect to $T$, for example, which play a key role in short- and long-read alignment~\cite{li2013aligning}.  In fact, with additional practical $O (n \log \sigma)$-bit data structures we can quickly list the starting positions of all the copies in $T$ of any substring $P [i..j]$ of $P$.

In the worst-case we cannot store a text of length $n$ over an alphabet of size $\sigma$ in fewer than $n \lg \sigma$ bits, and for a single genome using $2 n$ bits is reasonable. 
Due to the high speed and low cost of next-generation sequencing, however, we now have massive genomic databases to index, which are far more compressible than even what their high-order empirical entropies would indicate.

A consensus is emerging that if $T$ is such a text then compressed indexes for it should take space bounded in terms the number $r$ of runs in its Burrows-Wheeler Transform (\bwt), where a run is a maximal non-empty unary substring.  Within $O(r)$ space, we can efficiently support several powerful queries, but not yet computing \ms. Indeed, \ms{} can be easily computed with suffix trees, but suffix-tree functionality has been achieved only with $O (r \log (n / r))$-space data structures~\cite{gagie2020fully}, which is significantly larger than $O(r)$ both in theory and in practice~\cite{NO16}.

Bannai et al.\ recently presented their two-pass algorithm for quickly computing \ms\ using only an $O (r)$-space data structure during the first pass, from right to left in $O (m \log \log n)$ time, and then random access to $T$ during the second pass, from left to right.  We do not know how to support efficient random access to $T$ using only $O (r)$ space, however.

Once we have \ms, we can use the $r$-index \cite{gagie2020fully}, which also takes $O (r)$ words of space, to list the starting positions of all the copies in $T$ of any substring $P [i..j]$ of $P$, in $O (\log \log n)$ time per copy.  Specifically, we use Theorem~\ref{thm:JACM} below with Algorithm~\ref{algListCopies},
where
\begin{itemize}
\item $\phi (p) = \sa [\isa [p] - 1]$ (or \NULL\ if $\isa [p] = 0$),
\item $\phi^{-1} (p) = \sa [\isa [p] + 1]$ (or \NULL\ if $\isa [p] = n - 1$),
\item $\plcp [p] = \lcp [\isa [p]]$ (or $0$ if $\isa [p] = 0$),
\end{itemize}
and \sa, \isa, \lcp\ and \plcp\ are the suffix array, inverse suffix array, longest-common-prefix array and permuted longest-common-prefix array of $T$, respectively.

\begin{theorem}[Gagie, Navarro and Prezza, 2020]
\label{thm:JACM}
We can store $T$ in $O (r)$ space such that, given a text position~$p \in [0.. n - 1]$, we can compute $\phi (p)$, $\phi^{-1} (p)$ and $\plcp [p]$
in $O (\log \log n)$ time.
\end{theorem}

\begin{algorithm}[t]
	\DontPrintSemicolon
	\lIf{$\ms [i].\len < j - i + 1$}{\Return
}
$p \gets \ms [i].\pos$ \;
{\bf output} $p$ \;
\While{$\plcp [p] \geq j - i + 1$}{$p \gets \phi (p)$ \;
	{\bf output} $p$
}
$p \gets \phi^{-1} (\ms [i].\pos)$ \;
\While{$p \neq \NULL$ \textup{and} $\plcp [p] \geq j - i + 1$}{{\bf output} $p$ \;
	$p \gets \phi^{-1} (p)$
}

\caption{Lists the starting positions of all the copies of $P [i..j]$ in $T$
for given \ms{}, $i$ and $j$,
by using $\phi$, $\phi^{-1}$ and $\plcp$. }
\label{algListCopies}
\end{algorithm}

\subsection{Random Access in Compressed Space}
\label{subsec:random_access}

There are many data structures supporting logarithmic-time random access to compressed repetitive texts. For example, one can build a balanced grammar of size $O(z\log(n/z))$ that supports access to any symbol in time $O(\log(n/z))$ \cite{rytter2003application,charikar2005smallest},
where $z$ is the number of phrases in the LZ77 parse of $T$.  In practice, heuristics like \mbox{RePair}~\cite{LarssonM99} yield smaller balanced grammars.

Gagie et al.~\cite{gagie2019rpair} recently showed how to scale up RePair to handle genomic databases in reasonable time, using as a preprocessing step prefix-free parsing (PFP), a technique Boucher et al.~\cite{boucher2019prefix,kuhnle2020efficient} introduced to ease the construction of $\bwt$s of genomic database sequences.
PFP parses $T$ by passing a sliding window over it, inserting a phrase break when the Karp-Rabin hash of the contents of the window is 0 modulo a parameter.  PFP outputs a {\em dictionary} of phrases, and a {\em parse}: a sequence of dictionary symbols that when replaced by the corresponding phrases yields the original text~$T$. PFP produces mostly locally consistent parsings in practice, meaning that long repeated substrings in $T$ tend to be parsed roughly the same way, so the sum of the total lengths of the phrases in the dictionary and the number of phrases in the parse is usually significantly less than~$n$.

To get an SLP for $T$, Gagie et al.~\cite{gagie2019rpair} run RePair on the concatenation of the phrases in the dictionary, separated by unique symbols, to obtain an SLP containing a non-terminal for each phrase, whose expansion is that phrase.  Then, they run RePair on the parse, treating it as a string of phrase identifiers, to obtain an SLP whose terminals are the phrase identifiers.  By replacing the terminals in the latter SLP by the corresponding non-terminals in the former SLP, they obtain an SLP for $T$.  Since the dictionary and the parse are significantly smaller than $T$, running RePair on them is usually much faster and uses much less memory than running RePair (directly) on $T$, but the resulting SLP is usually only negligibly larger (and much smaller than Bannai et al.'s~\cite{bannai2020refining} data structure, anyway).

The na\"ive way to augment an SLP to support fast random access is to store with each non-terminal the size of its expansion.  Very recently Gagie et al.~\cite{gagie2020practical} gave a more space-efficient way to encode SLPs while still supporting fast random access. Rossi et al.~\cite{MONI} use this space-efficient SLP to support random access to $T$ in their implementation of Bannai et al.'s  algorithm.

\subsection{\lce\ Queries}
\label{subsec:lce}

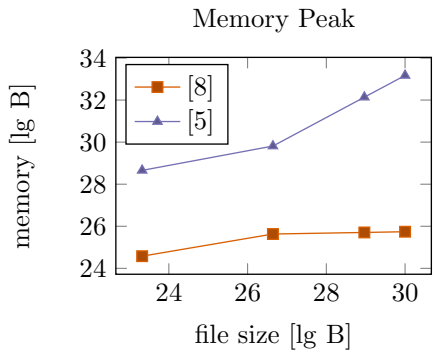
\begin{wrapfigure}{L}{0.4\textwidth}
\vspace{-1.5em}
\begin{tikzpicture}
\begin{axis}[
width=0.9\linewidth,
height=0.7\linewidth,
title={Memory Peak},
ylabel={memory [$\lg$ B]},
xlabel={file size [$\lg$ B]},
legend columns=1
]

\pgfplotsset{cycle list shift=1} \addplot coordinates{(23.32192809488736, 24.57817692965369) (26.643856189774727, 25.631540584259575) (28.96578428466209, 25.706076944475143) (30.0, 25.740399669191)};
\addlegendentry{\cite{gagie2020practical}};
\pgfplotsset{cycle list shift=1} \addplot coordinates{(23.32192809488736, 28.656964972669645) (26.643856189774727, 29.81406219114466) (28.96578428466209, 32.13576549046862) (30.0, 33.16995244525308)};
\addlegendentry{\cite{dinklage2020practical}};
 \end{axis}
\end{tikzpicture}
\caption{Memory needed for LCE queries on Chromosome 19 samples of various sizes (x-axis).}
\label{figLCEqueries}
\end{wrapfigure}
The longest common extension (\lce) query~\cite{jda/IlieNT10} asks, given two text positions $i$, $j$, for the length of the longest common prefix between $T[i..n-1]$ and $T[j..n-1]$. We make use of \lce{} queries to avoid the second pass in Bannai et al.'s algorithm~\cite{bannai2020refining}. 

Although there are many \lce\ data structures in the literature, we are not aware of any that in practice can handle genomic databases, which can range from tens of gigabytes to petabytes, while achieving compression comparable to Bannai et al.'s~\cite{bannai2020refining} or Gagie et al.'s~\cite{gagie2020practical} data structures and supporting reasonably fast queries.  For example, Dinklage et al.~\cite{dinklage2020practical} recently presented an \lce\ data structure based on sampling. Since its space usage grows linearly with the size of the dataset, this approach is impractical when dealing with genomic databases; see Figure~\ref{figLCEqueries} for a comparison of the memory usage between $\text{sss}_{256}$ of~\cite{dinklage2020practical} and the SLP representation of Gagie et al.~\cite{gagie2020practical}.

We now describe a practical algorithm for \lce\ queries that uses the same SLP compressed text representation~\cite{gagie2020practical} that Rossi et al.~\cite{MONI} used to random-access the input text. Because PFP produces mostly locally consistent parsings in practice, if we are extracting and comparing two suffixes of $T$ that have a long common prefix then usually we will quickly reach a phrase boundary simultaneously in both suffixes.  At that point we can start comparing the suffixes phrase by phrase instead of character by character, without expanding the non-terminals representing those phrases.  Once we reach phrases that do not match, we can expand their non-terminals and once again compare the suffixes character by character, knowing that we will find characters that do not match within those phrases.

We need not augment the SLP with information about which non-terminals' expansions are phrases: to extract and compare suffixes of $T$ starting at $T [i]$ and $T [j]$, conceptually we descend to the $i$th and $j$th leaves of the parse tree for $T$; starting from those leaves, we then perform synchronized traversals of the parse tree, moving from left to right until we simultaneously arrive at two leaves labelled with different characters; 
if during the traversals we simultaneously arrive at nodes labelled with the same non-terminal then we need not explore those nodes' subtrees --- whether or not that non-terminal's expansion is a phrase --- since they are guaranteed to be equal. This technique is described in detail with a similar parsing in Fischer et al.~\cite[Sect.~3.3]{fischer20deterministic}; see also Nishimoto et al.~\cite{NIIBT16}.

\section{Simplifying Bannai et al.'s Algorithm}
\label{sec:simplifying}

Bannai et al.\ developed their algorithm starting from Algorithm~\ref{algMatchingStatistics}, 
with which we can compute \ms\ in $m \log^{O (1)} n$ time with $O (n \log \sigma)$-bit data structures.  (The precise complexity depends on the auxiliary data structures used.)  In this algorithm, $\bwt.\rank_x (y)$ returns the number of copies of $x$ in $\bwt [0..y]$, $\bwt.\select_x (y)$ returns the position of the $y$th copy of $x$ in $\bwt$, and $\lf (x)$ returns the position in \bwt\ of the character that precedes $\bwt [x]$ in $T$.  We refer the reader to Navarro's text book~\cite{navarro2016compact} for descriptions of $O (n \log \sigma)$-bit data structures with which we can implement \rank, \select, \sa, \lf\ and \lce\ queries efficiently.  For example, we can implement \lce\ queries using a $(2 n + o (n))$-bit position-only range-minimum-query data structure over \lcp, and simulating \lcp\ using random access to \sa\ and \plcp.

\begin{algorithm}[t]
\DontPrintSemicolon
$q \gets \bwt.\select_{P [m - 1]} (1)$ \Comment*{Assume that $P[m-1]$ occurs in $T$} 
$\ms [m - 1] \gets (\pos : \sa [q] - 1$, \len: 1) \;
$q \gets \lf (q)$ \Comment*{Invariant: $T[\sa[q]] = P[m-1]$}
\For{$i = m - 2$ \textup{\bf down to} $0$}{\If{$\bwt [q] = P[i]$}{\label{lineBWTMatchPattern}
		$\ms [i] \gets (\pos : \ms[i + 1].\pos - 1, \len : \ms[i + 1].\len + 1)$ \label{lineBWTreduciblePosition} \;
        $q \gets \lf (q)$ \;
		}\Else{\label{lineBWTMismatchPattern}
        $c \gets \bwt.\rank_{P [i]} (q)$ \;
        $q' \gets \bwt.\select_{P [i]} (c)$ \;
        $q'' \gets \bwt.\select_{P [i]} (c + 1)$ \;
		$\ell' \gets \min (\ms [i + 1].\len, \lce (\sa [q'], \ms [i + 1].\pos))$ \label{lineLCEQ1} \;

		$\ell'' \gets \min (\ms [i + 1].\len, \lce (\sa [q''], \ms [i + 1].\pos))$ \label{lineLCEQ2} \;
		\If{$\ell' \geq \ell''$}{\label{lineSelectLargerLCE}
			$\ms [i] \gets (\pos : \sa [q'] - 1, \len :  \ell' + 1)$ \;
            $q \gets \lf (q')$ \;
		}\Else{$\ms [i] \gets (\pos : \sa [q''] - 1, \len : \ell'' + 1)$ \;
			$q \gets \lf (q'')$ \;
		}
	}
}

\caption{Computes \ms\ using $O (m)$ \rank, \select, \sa, \lf\ and \lce\ queries.  For simplicity we ignore the cases where $q$, $q'$ or $q''$ are undefined.}
\label{algMatchingStatistics}
\end{algorithm}

The crucial observation behind this algorithm is that when we know $\ms [i + 1]$ (both \pos\ and \len\ components):
\begin{itemize}
\item if $T [\ms [i + 1].\pos - 1] = P[i]$ then $\ms [i].\pos = \ms [i + 1].\pos - 1$ and $\ms [i].\len = \ms [i + 1].\len + 1$ (Line~\ref{lineBWTMatchPattern} in Algo.~\ref{algMatchingStatistics});
\item otherwise (Line~\ref{lineBWTMismatchPattern} in Algo.~\ref{algMatchingStatistics}), a copy of the longest prefix of $P [i..m - 1]$ that occurs in $T$ starts at either $T [p' - 1]$ or $T [p'' - 1]$, where $p'$ and $p''$ are the starting positions of 
the lexicographically preceding and succeeding suffixes of $T [\ms [i + 1].\pos..n - 1]$, respectively.
\end{itemize}
In the latter case, 
by knowing the lexicographic rank ($q$ in Algo.~\ref{algMatchingStatistics}) of $T [\ms [i+1].\pos..n - 1]$ among the suffixes of $T$,
we can tell whether to set $\ms [i].\pos$ to $p'-1$ or to $p''-1$ by comparing $\lce (p', \ms [i + 1].\pos)$ to $\lce (p'', \ms [i + 1].\pos)$ 
(Line~\ref{lineSelectLargerLCE} in Algorithm~\ref{algMatchingStatistics}).

Bannai et al.\ observed that most of the operations in this algorithm can be supported quickly using $O (r)$-space data structures.  For example, we need access to \sa\ entries only at positions corresponding to the starting or ending positions of runs in \bwt, and we can store those entries in $O (r)$ space.  The only exceptions are the \lce\ queries: the standard approach of solving \lce\ queries using range-minimum-query data structures over \lcp\ cannot be applied since we have no fast $O (r)$-space range-minimum-query data structures and, even if we did, we do not know how to support random access to \sa\ in order to use \plcp\ to simulate \lcp.

To avoid using \lce\ queries, Bannai et al.\ observed that during the execution of the algorithm it holds 
\begin{equation}\label{eq:lce}
\lce (p', \ms [i + 1].\pos) ~\geq~ \lce (p'', \ms [i + 1].\pos)
\end{equation}
if and only if there is a copy of $\min (\lcp [q' + 1], \cdots, \lcp [q''])$ in $\lcp [q + 1..q'']$.  Therefore, if we store the position of the first minimum in $\lcp [q' + 1..q'']$, for each choice of $q'$ and $q''$ as the ending position of a run in \bwt\ and the starting position of the next run of the same character, then we can compute the \pos\ components of the \ms\ array: $\ms [m-1].\pos, \ms [m-2].\pos, \ldots, \ms [0].\pos$ in this order. These values can be computed in overall $O (m \log \log n)$ time, since the cost is dominated by \rank\ and \select\ queries.

Without \lce\ queries it is not possible to compute the \len\ components together with the \pos\ components in an efficient way, so they are computed with a second, left to right, pass over the pattern $P$.  
Since $\ms [i + 1].\len \geq \ms [i].\len - 1$, once we know $\ms [i].\len$ we can find $\ms [i + 1].\len$ without looking at $T [\ms [i + 1].\pos..\ms [i + 1].\pos + \ms [i].\len - 2]$.  The total number of random accesses to $T$ we need forms a telescoping sum, 
\[
	\ms [0].\len + \sum_{i = 1}^{m - 1} (\ms [i].\len - \ms [i - 1].\len) + O (m),
\]
which collapses to $O (m)$.

Rossi et al.~\cite{MONI} described an implementation of the resulting two-pass algorithm providing also the missing step of the actual computation of the minima, called \emph{thresholds} in~\cite{bannai2020refining}, required to indirectly perform the comparison of Equation~\eqref{eq:lce}.

\newcommand{\MSFast}[1]{\ensuremath{\mathsf{msfast}_{\textup{#1}}}} \newcommand{\PhoniHeur}[1]{\ensuremath{\mathsf{PHONI}_{\textup{heur}}}}

\newcommand{\PhoniNaive}[1]{\ensuremath{\mathsf{PHONI}_{\textup{na\"ive}}}}
\newcommand{\PhoniStd}[1]{\ensuremath{\mathsf{PHONI}_{\textup{std}}}}

\section{Experiments}
\label{sec:experiments}

We compared the time and memory needed for building and querying \Moni{}, \Phoni{}, and Belazzougui et al.'s data structure~\cite{belazzougui2018fast} called \MSFast{} in the sequel.
In particular, we focus on three variants of \Phoni{} and two variants of \MSFast{}, which only differ in how the queries are performed:
\begin{itemize}
	\item \PhoniStd{} is what we have described in previous sections.
	\item \PhoniNaive{} performs \lce\ queries via character-by-character extraction, without skipping equal non-terminals as described in Subsection~\ref{subsec:lce}.
	\item \PhoniHeur{} changes the order of the execution of the LCE queries at Lines~\ref{lineLCEQ1} and~\ref{lineLCEQ2} of Algo.~\ref{algMatchingStatistics} favoring the 
		LCE query of $q_{\text{min}} \in \{q', q''\}$ with $|q - q_\text{min}| = \min(q-q', q''-q)$, and then omitting the second LCE query whenever the returned LCE length is at least $\ms[i+1].\Len$.
	\item \MSFast{} is the execution of Belazzougui et al.'s data structure with default parameters.
	\item \MSFast{+} is the execution of \MSFast{} with the parameters \texttt{-lazy\_wl 1 -double\_rank 1 -rank\_fail 1} leading to the fastest execution.
\end{itemize}
All the implementation is done in \CPlusPlus{17}. 
We ran our experiments on a machine with an Intel Xeon CPU E5-2620 and 64 GiB RAM running Ubuntu 16.04.7.  
We executed all programs single threaded.
We used the implementation \url{https://github.com/odenas/indexed_ms} and \url{https://github.com/maxrossi91/moni} for \MSFast{} and \Moni{}, respectively. 

We built the data structures for datasets consisting of chromosome 19s from human haplotypes, from 16 to 1000, 
and queried them with prefixes of 10 different chromosome 19s.  
Table~\ref{tablePatternCharateristics} gives some characteristics of the used data and the resulting matching statistics.  

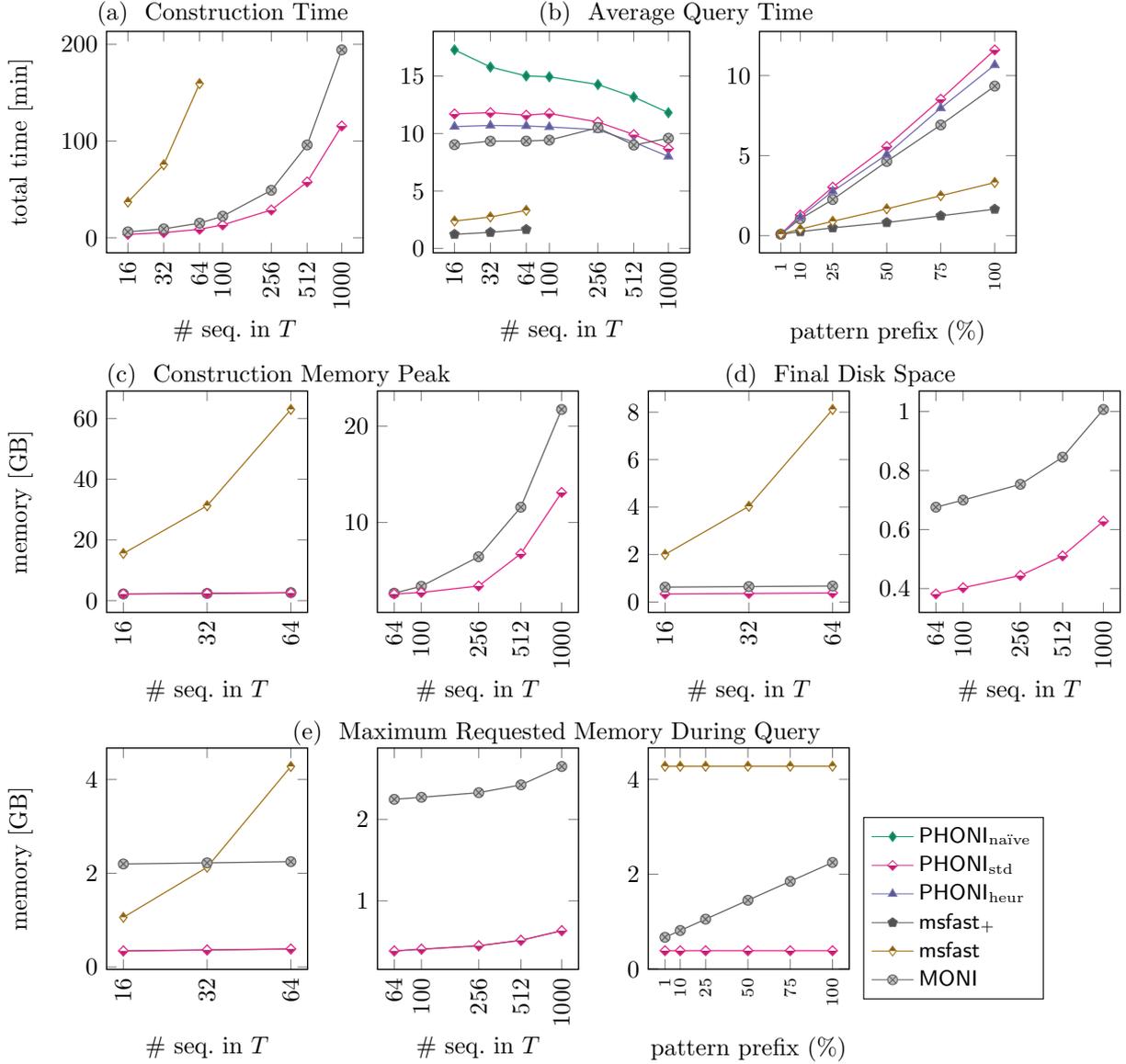
\begin{figure}[t!]
\usepgfplotslibrary{groupplots}

\newcommand*{\SamplesT}{\# seq.\ in $T$}
\newcommand*{\SamplesP}{pattern prefix (\%)}

\pgfplotsset{myPlot/.style={squeezedPlot,
xtick=data,
log ticks with fixed point,
xticklabel style={rotate=90, anchor=east},
x label style={at={(axis description cs:0.5,-0.1)},anchor=north},
log basis x=2,
/pgf/number format/1000 sep={},
},
patternScale/.style={xlabel={\SamplesP{}},
every x tick label/.append style={font=\scriptsize},
}
}

\begin{minipage}{0.4\linewidth}
	\centering
\myLabel{fig:construction_time}{(a)} Construction Time
\end{minipage}
\begin{minipage}{0.4\linewidth}
	\centering
	\myLabel{fig:query_time}{(b)} Average Query Time
\end{minipage}
\hfill

\begin{tikzpicture}
\begin{groupplot}[group style = {group size = 2 by 1},
xlabel={\SamplesT{}},
ylabel={total time [min]},
xmode=log, myPlot,
width=0.33\linewidth,
group style={group name=my plots,
        group size=4 by 1,
        ylabels at=edge left
    },
]

\nextgroupplot{}
\pgfplotsset{cycle list shift=3} \addplot coordinates{(16, 3.582166666666667) (32, 5.371666666666666) (64, 8.8285) (100, 13.347) (256, 28.753833333333333) (512, 57.72366666666667) (1000, 115.53883333333336)};
\addlegendentry{\PhoniStd{}};
\pgfplotsset{cycle list shift=5} \addplot coordinates{(16, 36.8645) (32, 75.55316666666666) (64, 159.33950000000002)};
\addlegendentry{\MSFast{}};
\pgfplotsset{cycle list shift=5} \addplot coordinates{(16, 6.145333333333334) (32, 9.221499999999999) (64, 15.161333333333335) (100, 22.324333333333332) (256, 49.113166666666665) (512, 95.9765) (1000, 194.26683333333332)};
\addlegendentry{\Moni{}};
 \legend{}

\nextgroupplot[appendLegend,
legend to name={legBenchmark},
]{}
\pgfplotsset{cycle list shift=8} \addplot coordinates{(16, 17.283516666666667) (32, 15.789583333333333) (64, 15.013449999999999) (100, 14.93795) (256, 14.262749999999999) (512, 13.18095) (1000, 11.812033333333334)};
\addlegendentry{\PhoniNaive{}};
\pgfplotsset{cycle list shift=2} \addplot coordinates{(16, 11.709200000000001) (32, 11.820799999999998) (64, 11.601799999999999) (100, 11.74585) (256, 11.002233333333333) (512, 9.909766666666666) (1000, 8.699216666666667)};
\addlegendentry{\PhoniStd{}};
\pgfplotsset{cycle list shift=0} \addplot coordinates{(16, 10.602549999999999) (32, 10.705233333333334) (64, 10.662983333333333) (100, 10.57225) (256, 10.3316) (512, 9.254566666666665) (1000, 8.004666666666667)};
\addlegendentry{\PhoniHeur{}};
\pgfplotsset{cycle list shift=12} \addplot coordinates{(16, 1.2302333333333333) (32, 1.4056333333333333) (64, 1.6588666666666667)};
\addlegendentry{\MSFast{+}};
\pgfplotsset{cycle list shift=2} \addplot coordinates{(16, 2.3893999999999997) (32, 2.741966666666667) (64, 3.3211500000000003)};
\addlegendentry{\MSFast{}};
\pgfplotsset{cycle list shift=2} \addplot coordinates{(16, 9.031283333333333) (32, 9.339833333333333) (64, 9.344033333333334) (100, 9.429850000000002) (256, 10.508466666666667) (512, 8.997183333333334) (1000, 9.592966666666666)};
\addlegendentry{\Moni{}};

\nextgroupplot[patternScale,xmode=normal]
\pgfplotsset{cycle list shift=7} \addplot coordinates{(1, 0.09123333333333333) (10, 1.0533833333333333) (25, 2.2605666666666666) (50, 4.63565) (75, 6.917416666666666) (100, 9.344033333333334)};
\addlegendentry{\Moni{}};
\pgfplotsset{cycle list shift=2} \addplot coordinates{(1, 0.08941666666666666) (10, 1.29665) (25, 3.0342666666666664) (50, 5.570916666666667) (75, 8.524416666666665) (100, 11.601799999999999)};
\addlegendentry{\PhoniStd{}};
\pgfplotsset{cycle list shift=0} \addplot coordinates{(1, 0.08246666666666666) (10, 1.1717166666666665) (25, 2.7718) (50, 5.066983333333334) (75, 7.9701) (100, 10.662983333333333)};
\addlegendentry{\PhoniHeur{}};
\pgfplotsset{cycle list shift=12} \addplot coordinates{(1, 0.09036666666666665) (10, 0.2513666666666667) (25, 0.4913666666666667) (50, 0.8229833333333334) (75, 1.24495) (100, 1.6588666666666667)};
\addlegendentry{\MSFast{+}};
\pgfplotsset{cycle list shift=2} \addplot coordinates{(1, 0.09776666666666667) (10, 0.4147666666666666) (25, 0.9022666666666667) (50, 1.6790166666666666) (75, 2.4973666666666663) (100, 3.3211500000000003)};
\addlegendentry{\MSFast{}};
 \legend{}
\end{groupplot}
\end{tikzpicture}

\begin{minipage}{0.5\linewidth}
	\centering
	\myLabel{fig:construction_memory}{(c)} Construction Memory Peak
\end{minipage}
\begin{minipage}{0.5\linewidth}
	\centering
	\myLabel{fig:final_space}{(d)} Final Disk Space
\end{minipage}

\begin{tikzpicture}

	\begin{groupplot}[group style = {group size = 2 by 1},
xlabel={\SamplesT{}},
ylabel={memory [GB]},
xmode=log, myPlot,
width=0.28\linewidth,
group style={group name=my plots,
        group size=4 by 1,
        ylabels at=edge left
    },
]

\nextgroupplot{}
\pgfplotsset{cycle list shift=7} \addplot coordinates{(16, 2.200952) (32, 2.37626) (64, 2.642348)};
\addlegendentry{\Moni{}};
\pgfplotsset{cycle list shift=5} \addplot coordinates{(16, 15.526656) (32, 31.256884) (64, 62.970752)};
\addlegendentry{\MSFast{}};
\pgfplotsset{cycle list shift=1} \addplot coordinates{(16, 2.200952) (32, 2.37626) (64, 2.56276)};
\addlegendentry{\PhoniStd{}};
 \legend{}

\nextgroupplot{}
\pgfplotsset{cycle list shift=7} \addplot coordinates{(64, 2.642348) (100, 3.360324) (256, 6.433532) (512, 11.582668) (1000, 21.742268)};
\addlegendentry{\Moni{}};
\pgfplotsset{cycle list shift=2} \addplot coordinates{(64, 2.56276) (100, 2.717076) (256, 3.390476) (512, 6.744652) (1000, 13.11544)};
\addlegendentry{\PhoniStd{}};
 \legend{}

\nextgroupplot{}
\pgfplotsset{cycle list shift=3} \addplot coordinates{(16, 0.340529738) (32, 0.360752975) (64, 0.381887039)};
\addlegendentry{\PhoniStd{}};
\pgfplotsset{cycle list shift=6} \addplot coordinates{(16, 0.628424281) (32, 0.651147031) (64, 0.675799778)};
\addlegendentry{\Moni{}};
\pgfplotsset{cycle list shift=4} \addplot coordinates{(16, 2.00484303) (32, 4.0324039) (64, 8.110526346)};
\addlegendentry{\MSFast{}};
 \legend{}

\nextgroupplot{}
\pgfplotsset{cycle list shift=3} \addplot coordinates{(64, 0.381887039) (100, 0.403070907) (256, 0.444439602) (512, 0.510939309) (1000, 0.628279006)};
\addlegendentry{\PhoniStd{}};
\pgfplotsset{cycle list shift=6} \addplot coordinates{(64, 0.675799778) (100, 0.699879003) (256, 0.75325257) (512, 0.845318687) (1000, 1.006911999)};
\addlegendentry{\Moni{}};
 \legend{}
\end{groupplot}
\end{tikzpicture}

\begin{minipage}{\linewidth}
	\centering
	{\myLabel{fig:query_space}{(e)} Maximum Requested Memory During Query}
\end{minipage}

\begin{adjustbox}{valign=c}
\begin{tikzpicture}
	\begin{groupplot}[group style = {group size = 2 by 1},
xlabel={\SamplesT{}},
ylabel={memory [GB]},
xmode=log, myPlot,
width=0.28\linewidth,
group style={group name=my plots,
        group size=4 by 1,
        ylabels at=edge left
    },
]

\nextgroupplot{}
\pgfplotsset{cycle list shift=8} \addplot coordinates{(16, 0.344348329) (32, 0.364613849) (64, 0.385961697)};
\addlegendentry{\PhoniNaive{}};
\pgfplotsset{cycle list shift=2} \addplot coordinates{(16, 0.344348329) (32, 0.364613849) (64, 0.385961697)};
\addlegendentry{\PhoniStd{}};
\pgfplotsset{cycle list shift=4} \addplot coordinates{(16, 1.059507897) (32, 2.127965481) (64, 4.276383577)};
\addlegendentry{\MSFast{}};
\pgfplotsset{cycle list shift=4} \addplot coordinates{(16, 2.197996325) (32, 2.221607917) (64, 2.247123085)};
\addlegendentry{\Moni{}};
 \legend{}

\nextgroupplot{}
\pgfplotsset{cycle list shift=8} \addplot coordinates{(64, 0.385961697) (100, 0.407497398) (256, 0.449677398) (512, 0.516583902) (1000, 0.634604915)};
\addlegendentry{\PhoniNaive{}};
\pgfplotsset{cycle list shift=2} \addplot coordinates{(64, 0.385961697) (100, 0.407497398) (256, 0.449677398) (512, 0.516583902) (1000, 0.634603379)};
\addlegendentry{\PhoniStd{}};
\pgfplotsset{cycle list shift=5} \addplot coordinates{(64, 2.247123085) (100, 2.27221928) (256, 2.329129584) (512, 2.424938072) (1000, 2.651001824)};
\addlegendentry{\Moni{}};
 \legend{}

\nextgroupplot[patternScale,xmode=normal]{}
\pgfplotsset{cycle list shift=7} \addplot coordinates{(1, 0.668618073) (10, 0.812119512) (25, 1.051288555) (50, 1.449903621) (75, 1.848518685) (100, 2.247123085)};
\addlegendentry{\Moni{}};
\pgfplotsset{cycle list shift=2} \addplot coordinates{(1, 0.385961189) (10, 0.385961705) (25, 0.385961705) (50, 0.385961705) (75, 0.385961705) (100, 0.385961697)};
\addlegendentry{\PhoniStd{}};
\pgfplotsset{cycle list shift=4} \addplot coordinates{(1, 4.275055551) (10, 4.275161853) (25, 4.275339015) (50, 4.275634287) (75, 4.275929559) (100, 4.276383577)};
\addlegendentry{\MSFast{}};
 \legend{}

\end{groupplot}
\end{tikzpicture}
\end{adjustbox}
\begin{adjustbox}{valign=c}
\ref{legBenchmark}
\end{adjustbox}

\label{figBenchmark}
\caption{The time {\bf \ref{fig:construction_time}} and memory {\bf \ref{fig:construction_memory}} needed to build \PhoniNaive{}, \PhoniStd{}, \Moni{} and \MSFast{}, and their final sizes {\bf \ref{fig:final_space}}; the query times {\bf \ref{fig:query_time}} as a function of dataset size and pattern length; and the memory used for queries {\bf \ref{fig:query_space}}.  The dataset used for testing prefixes of queries consisted of 64 chromosome 19s.
}
\end{figure}

Figure~\ref{figBenchmark}\ref{fig:construction_time} shows the construction times and~\ref{figBenchmark}\ref{fig:query_time} the query times, with the average time to compute \ms\ for a whole chromosome 19 depending on the size of the dataset on the left, and to compute \ms\ for a prefix of a chromosome 19 on the right.  
We stopped constructions that took more than 200 minutes.
We omit the variations of \Phoni{} and \MSFast{} in Figures~\ref{figBenchmark}\ref{fig:construction_time} and~\ref{figBenchmark}\ref{fig:construction_memory} because their constructions are the same as their respective standard variants.  
Additionally, their memory consumption is at most slightly different to their standard variants, such that we also omit them in Figure~\ref{figBenchmark}\ref{fig:query_space}.

We can clearly observe that \PhoniStd{} is faster than \PhoniNaive{} while being slightly inferior to \PhoniHeur{}.
\PhoniStd{} is faster to build than \Moni{} because it does not need the thresholds, but it is slower at answering queries for relatively small $|T|$ (although it speeds up as the dataset grows, because the fraction of the time we use \lf\ mappings increases).

Figures~\ref{figBenchmark}\ref{fig:construction_memory} and~\ref{fig:final_space} show the peak RAM (the maximum resident set size) used to build the data structures and their final sizes.  
\PhoniStd{} also takes less memory that \Moni{} because, again, it does not need thresholds:
While the data points of \PhoniStd{} and \Moni{} seem to overlap in the left plot of Figures~\ref{figBenchmark}\ref{fig:construction_memory}, the right plot makes the gap consisting of the thresholds clearly visible for a larger number of indexed sequences. 
Finally, Fig.~\ref{figBenchmark}\ref{fig:query_space} shows the maximum amount of memory requested for allocation during a query (including the indexing data structures).

For the larger datasets, \Moni{} and \PhoniStd{} seem to be the most practical solutions, with \Moni{} being faster for small datasets but somewhat harder to build and needing more RAM when computing the matching statistics for long patterns.  As stated in Section~\ref{sec:introduction}, for the full version of this paper we will test the maximum latency and the RAM usage when the solutions process queries in parallel.

\paragraph{Construction of \Phoni{}.}
To construct \Phoni{}, we used Rossi et al.'s construction algorithm for \Moni{} to build the RLBWT and the SLP\@.
The indexing data structures of \Moni{} assume multiple sequences whose characters are drawn from the byte alphabet.
We omit the construction of the thresholds from the index, as we do not need them.

\paragraph{\MSFast{}.}
Regarding times, we can see that the construction and query times grow linearly with the sample sizes.
While its construction is among one of the slowest, it is faster than \Phoni{} when it comes to queries.
However, due to the linear dependency on the sample size, 
we expect that this solution eventually becomes slower than \Phoni{} for larger sample sizes.
For the space, we observe that the construction of \MSFast{} has a huge memory footprint. 
Its memory requirement already reached the maximum available memory of our machine at 64 samples. 
Due to this physical restriction, there are no evaluations of \MSFast{} for larger sample sets.

\paragraph{\Moni{}.}
Regarding the measured times (Fig.~\ref{figBenchmark}\ref{fig:construction_time} and~\ref{fig:query_time}), 
we observe that \Moni{} with the additional need of the thresholds is slower during the construction than $\Phoni{1}$,
while excelling at the queries.
However, for larger sample sizes, the threshold computation take a considerable amount of time while \Phoni{}'s query times become faster as the number of reducible positions (cf.~\ref{tablePatternCharateristics}) increases,
making the time-expensive LCE queries less frequent.
At \num{1000} samples, the overall times are competitive, and we expect to see that \Phoni{} will take the lead with even more sequences.
While the difference in the construction is only the thresholds computation, 
\Phoni{} and \Moni{} largely differ in the needed memory requirements when it comes to queries.
Here, \Moni{} needs to additionally keep the patterns, the thresholds, and the \Pos{} components of the matching statistics in memory,
causing a large memory footprint observable in Fig.~\ref{figBenchmark}\ref{fig:query_space}.
Unlike \Moni{}, \Phoni{} can stream both the input patterns and the output, making it the most favorable option
when memory is the computational limitation.

\begin{table}[t]
		\begin{minipage}{0.43\linewidth}
		\begin{tabular}{*{5}{r}}
			\toprule
			\# & \multicolumn{1}{c}{$|T|$} & \multicolumn{1}{c}{$|$SLP$|$} & \multicolumn{1}{c}{$r$} & \lf \%\ \\
			   & \multicolumn{1}{c}{\small[GB]} & \multicolumn{1}{c}{\small[MB]} & \multicolumn{1}{c}{\small[M]} &
			\\\midrule
\num{16} & \num{0.96} & \num{36.11} & \num{32.40} & \num{78.88} \\
\num{32} & \num{1.92} & \num{37.86} & \num{32.83} & \num{79.11} \\
\num{64} & \num{3.85} & \num{39.49} & \num{33.34} & \num{79.36} \\
\num{100} & \num{6.01} & \num{41.02} & \num{33.78} & \num{79.56} \\
\num{256} & \num{15.39} & \num{47.38} & \num{35.62} & \num{80.34} \\
\num{512} & \num{30.78} & \num{57.98} & \num{39.24} & \num{81.96} \\
\num{1000} & \num{60.11} & \num{80.64} & \num{45.93} & \num{84.61} \\
 			\bottomrule
		\end{tabular}
\end{minipage}
\hfill
		\begin{minipage}{0.52\linewidth}
	\caption{Characteristics of the obtained matching statistics. 
		The column \# is the number of chr19 sequences stored in $T$, 
		$|$SLP$|$ the sizes of \PhoniStd{}'s SLP grammar,
		and $r$ the runs in \bwt{} (in million).
		The column \emph{\lf \% } is the percentage of how often Line~\ref{lineBWTMatchPattern} in Algo.~\ref{algMatchingStatistics} is true.
	This percentage increases with the number of samples since it becomes likelier for matches the longer the indexed text becomes.
		The average and maximum value of \Len{} in \ms{} are roughly \num{81558} and \num{3100685} for all instances.
}
		\end{minipage}
	\label{tablePatternCharateristics}
\end{table}

\bibliographystyle{abbrv}

\end{document}